\documentclass[nofootinbib,twocolumn,showpacs,preprintnumbers,superscriptaddress,prb,floatfix,aps,10pt]{revtex4-1}
\usepackage[printwatermark]{xwatermark}
\usepackage{verbatim} % for block comments
\usepackage{amsmath,amssymb}
\usepackage{graphicx,textcomp}
\usepackage{epstopdf}
\usepackage{multirow}
\usepackage{xfrac}
\usepackage[version=3]{mhchem}
\usepackage{siunitx}
\usepackage{xcolor}

\newcommand{\ts}[1]{\textsubscript{#1}}

\renewcommand{\vec}[1]{\ensuremath{\mathbf{#1}}}
\newcommand{\pvec}[1]{\ensuremath{\mathbf{#1}}}
\def\-{\raisebox{.75pt}{-}} % adjust length of minus sign

\begin{document}

%\newwatermark[allpages,color=red!50,angle=45,scale=2,xpos=0,ypos=0]{DO NOT DISTRIBUTE}

\title{Discovery of Ordered Vortex Phase \\ in Multiferroic Oxide Superlattices}

\author{Antonio B. Mei} \email{amei2@illinois.edu} \affiliation{Department of Materials Science and Engineering, Cornell University, Ithaca, NY, 14853, USA}
%\author{David Muller} \affiliation{Department of Materials Science and Engineering, Cornell University, Ithaca, NY, 14853, USA} 
\author{Ramamoorthy Ramesh} \affiliation{Department of Materials Science and Engineering, University of California, Berkeley, California 94720, USA} \affiliation{Materials Sciences Division, Lawrence Berkeley National Laboratory, Berkeley, California 94720, USA} \affiliation{Department of Physics, University of California, Berkeley, California 94720, USA}
\author{Darrell G. Schlom} \affiliation{Department of Materials Science and Engineering, Cornell University, Ithaca, NY, 14853, USA} \affiliation{Kavli Institute at Cornell for Nanoscale Science, Ithaca, NY, 14853, USA}

\maketitle

%\section{Introduction}

\textbf{
Ferroics, characterized by a broken symmetry state with nonzero elastic, polar, or magnetic order parameters $\vec{u}$, are recognized platforms for staging and manipulating topologically-protected structures\cite{Privratska:1999id,Thomas:2007jl,Seidel:2009hs,Balke:2011eb,Muhlbauer:2009bc,Yu:2011hr} as well as for detecting unconventional topological phenomena.\cite{Chambers:1960bx} The unrealized possibility of producing ordered topological phases in magnetoelectric multiferroics, exhibiting coupled magnetic and polar order parameters,\cite{Rovillain:2010kj} is anticipated to engender novel functionality and open avenues for manipulating topological features. Here, we report the discovery of an ordered $\pi_1$-$S_\infty$ vortex phase within single-phase magnetoelectric multiferroic BiFeO\ts{3}. The phase, characterized by positive topological charge and chiral staggering, is realized in coherent TbScO\ts{3} and BiFeO\ts{3} superlattices and established via the combination of direct- and Fourier-space analyses. Observed order-parameter morphologies are reproduced with a field model describing the local order-parameter stiffness and competing non-local dipole-dipole interactions. Anisotropies canting the order parameter towards $\left<100\right>$ suppress chiral staggering and produced a competing $\pi_1$-$C_{\infty v}$ vortex phase in which cores are centered.
}

\begin{figure}[!h]
\includegraphics[width=0.47\textwidth]{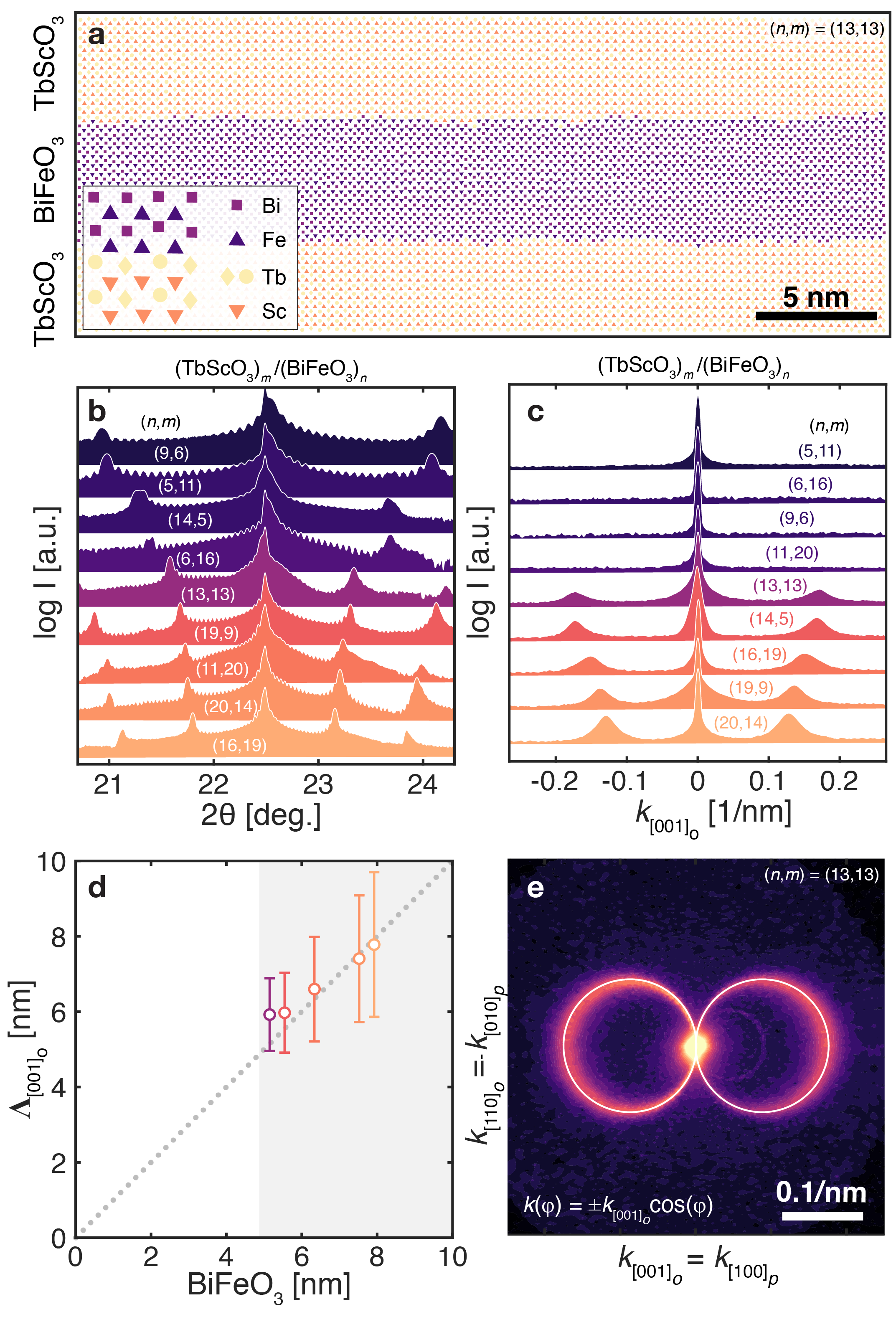}
\caption{\label{fig:XRD}
\textbf{Structural characterization of coherent (TbScO\ts{3})$_{m}$/(BiFeO\ts{3})$_{n}$ superlattices.}
(a) Classification of STEM atomic columns, collected down the $[010]_p$ zone axis, by an artificial neural network demonstrates sharp superlattice interfaces with negligible interlayer diffusion.
(b) XRD $\theta$-$2\theta$ scans near the fundamental $001_p$ film reflections exhibit thickness oscillations corroborating atomically abrupt interfaces. 
(c) In-plane XRD scans as a function of reciprocal distance $k_{[100]_p}$ from $001_p$ show ordered superstructure reflections in samples with $n \geq 13$ unit cells.
(d) Superstructure periodicities $\Lambda_{[100]_p}$ as a function of BiFeO\ts{3} layer thicknesses $n$.
(e) Representative in-plane RSM showing the azimuthal dependence of superstructure reflections. 
Scans in (b) and (c) are sorted by $n+m$ bilayer and $n$ BiFeO\ts{3} thicknesses, respectively, and offset for clarity. 
}
\end{figure}

%\section{Prerequisite for forming vortices}

When ferroic symmetries are broken, the order parameter $\vec{u}(\vec{r})$ adopts many possible degenerate ground states.\cite{Chaikin:2000td} Within the locality of a casually connected domain $\Omega$ the order parameter will be uniform ($\vec{u}(\vec{r}) = \left<\vec{u}\right>$ $\forall$ $\vec{r}$ $\subset \Omega$). However, between $\Omega$, structures may emerge as topologically-protected relics of the bygone global symmetries.\cite{Zurek:1985ko} The structures that form depend on the topology of $\mathcal{M}$ the manifold spanned by the degenerate equilibrium states.\cite{Kibble:1976sj} For vortices to persist in the broken symmetry state, loops must exist in $\mathcal{M}$ which are not continuously deformable into other loops, i.e. the fundamental group $\pi_1$ must be nontrivial. The simplest manifold fulfilling this homotopic condition is that described by the one-dimensional special unitary abelian Lie group $U(1)$.\cite{Kosterlitz:1973fc} In this case, the order parameter lies on a degenerate manifold isotropic to continuous rotations and is completely specified by a phase angle. While full rotations in $\mathcal{M}$ restore the phase angle of the order parameter, the contours traced by the rotations are dissimilar and non-homotopic, resulting in a topological charge $q$ being acquired with every winding. % For, the fundamental group of is isomorphic with group of integers under addition, $\pi_1 = \mathbb{Z}$.

The magnetoelectric multiferroic BiFeO\ts{3} crystalizes in the ideal perovskite structure (space group $Pm\bar{3}m$ $O^1_h$) above 1200 K.\cite{Catalan:2009ca} Upon cooling, the effective order-parameter potential-energy landscape undergoes a series a spontaneous symmetry breaking phase transitions which culminate in a spin-canted G-type antiferromagnetic magnetoelectric ground state ($R3c$ $C^6_{3v}$) below $650$ K.\cite{Kornev:2007jr} The descent in symmetry engenders eight degenerate ferroelectric variants for which order parameters $\vec{u}$ point along $\left<111\right>_p$  directions (the $p$ subscript indicates pseudocubic indices).\cite{Balke:2009fj} Strong crystalline anisotropy, locking $\vec{u}$ to the discrete subspace $\mathcal{M}$ spanned by $O_h$ symmetries, represent a major challenge\cite{Hong:2017tw} to decoupling\cite{Muhlbauer:2009bc} the order parameter from the underlying atomic structure and stabilizing vortices in BiFeO\ts{3}. 

To realize vortices in BiFeO\ts{3}, we engineer the effective order-parameter potential-energy landscape by employing heteroepitaxial templating to isolate planar symmetries and dimensional confinement to restore continuous $U(1)$ rotational symmetries. This is physically accomplished by forming a coherent superlattice consisting of alternating layers of multiferroic BiFeO\ts{3}$(001)_p$ and dielectric TbScO\ts{3}$(110)_o$ (the $o$ subscript denotes orthorhombic indices) on symmetry-breaking TbScO\ts{3}$(110)_o$ substrates with rectangular surface nets. We denote these superlattices by (TbScO\ts{3})$_m$/(BiFeO\ts{3})$_n$, in which $n$ and $m$ refers to the thickness, in unit cells, of the corresponding layers. %The $(n,m)$ parameter space is sampled between $5 \leq n,m \leq 20$ using an inscribed rotated central composite design.\cite{Myers:2014va}

%\section{XRD results}

The structural perfection of the superlattices are investigated using scanning transmission electron microscopy (STEM) and x-ray diffraction (XRD). In lattice-resolution STEM images, atomic columns exhibit signature intensity modulations reflecting their chemical constituents and local coordination environment. Fig. \ref{fig:XRD}(a) shows the result of decoding the embedded chemical and structural information by training an artificial neural network to analyze the contrast of columns projected along the $[010]_p$ zone axis of a (TbScO\ts{3})$_{13}$/(BiFeO\ts{3})$_{13}$ superlattice. Five categorically distinct columns are identified, corresponding to bismuth, iron, scandium, and two terbium variants -- the two variants result from orthorhombic dimerization. The orderly arrangement of the columns evince coherent superlattice of high structural perfection with abrupt interfaces.

XRD $\theta$-$2\theta$ scans collected near $110_o$ substrate reflections are plotted in Fig. \ref{fig:XRD}(b) as a function of bilayer thickness $n+m$. The scans are characterized by superlattice reflections positioned symmetrically around $001_p$ fundamental peaks. In addition, thickness oscillations corroborating atomically smooth interfaces are observed in all samples. Diffracted intensities measured as a function of distance along $\hat{x} = [100]_p = [001]_o$ from the symmetric $001_p$ reflections are shown in Fig. \ref{fig:XRD}(c). Superlattices comprised of BiFeO\ts{3} layers with $n < 13$ unit cells exhibit only fundamental reflections. In contrast, superlattices with $n \geq 13$ unit cells display additional reflections at $k_{[100]_p}(n)$ indicative of in-plane ordering. Superstructure periodicities $\Lambda_{[100]_p}(n) = 1/k_{[100]_p}(n)$,  Fig. \ref{fig:XRD}(d), are found to be independent of $m$ the dielectric layer thickness and increase linearly with $n$ the multiferroic layer thickness with a slope of unity. The linear scaling contrasts against the $\Lambda \propto \sqrt{n}$ behavior\cite{Kittel:1946hg} observed for thick single-component BiFeO\ts{3} layers\cite{Catalan:2008kj} and is consistent with other superlattice systems\cite{Tang:2015gw} displaying continuously rotating order parameters.

The evolution of superstructure reflections with azimuthal orientation is investigated using XRD in-plane reciprocal-space maps (RSM). A representative map, obtained for (TbScO\ts{3})$_{13}$/(BiFeO\ts{3})$_{13}$, is presented in Fig. \ref{fig:XRD}(e). With the fundamental reflection positioned at the center of the plot, in-plane ordering produces intensity maxima which trace a figure-eight pattern. The symmetry of the pattern is defined by two orthogonal mirror planes, $[100]_p$ and $[010]_p$ --- the absence of four-fold symmetry is a consequence of templating from the underlying orthorhombic TbScO\ts{3} substrate.\cite{Uecker:2008dr} Parameterizing the intensity maxima as a function of azimuthal angle $\phi$ through $k(\phi) = \pm k_{[100]_p} \cos(\phi)$ demonstrates that the superstructure is a quasi-one-dimensional array with a periodicity of $\Lambda_{[100]_p}$ along $\psi=0$. In the limit of $\phi = \pm \pi$ (along $\pm\hat{y} = \pm[010]_p = \pm[1\bar{1}0]_o$), the periodicity asymptotically approaches infinity. 

%\section{TEM results}

\begin{figure}[!hbt]
\includegraphics[width=0.47\textwidth]{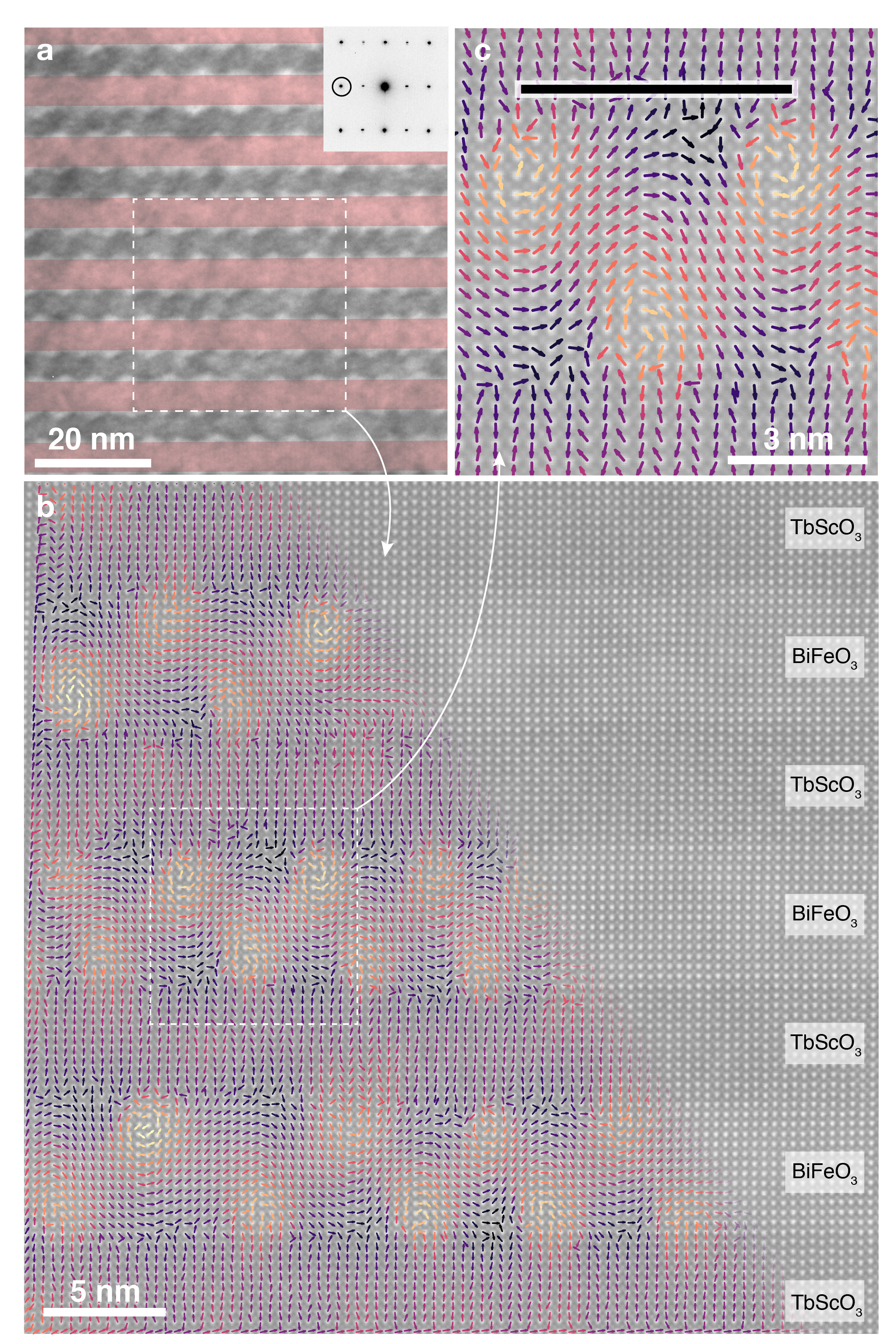}
\caption{\label{fig:TEM} 
\textbf{Observation of an ordered vortex phase in multiferroic BiFeO\ts{3}.}
(a) DF-TEM image collected near the $[010]_p$ zone axis of a (TbScO\ts{3})$_{13}$/(BiFeO\ts{3})$_{13}$ superlattice with $\bar{1}00_p$ excited under two-beam condition. Zone-axis SAED with the  $\bar{1}00_p$ reflection encircled (inset). The undulating pattern in BiFeO\ts{3} corroborates a chirally-staggered $\pi_1$-$S_\infty$ ordered vortex phase.
(b) STEM image overlaid with $\vec{u}(\vec{r})$ colorized based on local topological charge $q(\vec{r})$ .
(c) Enlarged region highlighting a continuously rotating order parameter $\vec{u}(\vec{r})$. The periodicity of the vortex phase obtained from STEM is consistent with that determined via XRD (indicated by the black horizontal bar).
Order-parameter fields in (b) and (c) are colorize based on topological charge $q(\vec{r})$ with bright areas representing positive $q(\vec{r})$.
}
\end{figure}

Direct-space images of the ordered superstructures are obtained using cross-sectional dark-field transmission electron microscopy (DF-TEM). Fig. \ref{fig:TEM}(a) is a micrograph of the same (TbScO\ts{3})$_{13}$/(BiFeO\ts{3})$_{13}$ superlattice, acquired near the $\hat{y} = [010]_p = [1\bar{1}0]_o$ zone axis. A corresponding zone-axis selected-area electron diffraction (SAED) pattern is provided in the insert; half order peaks in the in-plane direction are due to the orthorhombic distortion of TbScO\ts{3} [see Fig. \ref{fig:XRD}(a)]. With the objective aperture selecting $\vec{g} = \bar{1}00_p$ excited under two-beam condition, regions where $\vec{u}(\vec{r})$ and $\vec{g}$ are aligned appear bright, while areas where $\vec{u}(\vec{r})$ and $\vec{g}$ are anti-aligned appear dark.\cite{Williams:2016ju,Aoyagi:2011kp} Alternating layers corresponding to BiFeO\ts{3} and TbScO\ts{3} are discernible. In addition, strong contrast variations in the form of undulating patterns are present within BiFeO\ts{3}. The undulations develop from a periodic reversal of the in-plane component of the order parameter in the bright triangular wedges located near the dielectric interface and is consistent with an ordered vortex phase in which cores are positioned at the apex of each triangle.

Local order parameter variations are resolved with atomic precision by determining atomic displacement fields $\vec{u}(\vec{r})$ with subpixel accuracy from STEM  atomic column positions. Because of the bijective relationship between ferroelastic displacements and spontaneous polarizations, the position-dependent polarization field in ferroelectrics is uniquely determined by $\vec{u}(\vec{r})$.\cite{Balke:2009fj}  Fig. \ref{fig:TEM}(b) is a typical lattice-resolution STEM image overlaid with displacement fields $\vec{u}(\vec{r})$; an enlarged region is shown in Fig. \ref{fig:TEM}(c). Vector fields in Fig. \ref{fig:TEM}(b) and \ref{fig:TEM}(c) are colorized based on the local topological charge field:\cite{Fradkin:2013wwa}
\begin{equation}
\label{eq:q}
	q(\vec{r}) = \frac{1}{2\pi} \iint_S [ \partial_x \vec{u}(\vec{r}-\pvec{r}') \times \partial_z \vec{u}(\vec{r}-\pvec{r}') ] d^2\pvec{r}',
\end{equation}
which is computed by integrating the partials of the vector field $\partial_{\hat{e}}\vec{u}(\vec{r})$ with respect to $\hat{e}$ over vortex-containing surfaces $S$.
Within orthorhombic TbScO\ts{3}, $\vec{u}(\vec{r})$ is displaced along opposite out-of-plane directions in neighboring unit cells as a result of nonpolar dimerization [see Fig. \ref{fig:XRD}(a)]. Within BiFeO\ts{3}, an elaborate $\vec{u}(\vec{r})$ displacement pattern is observed. A key distinguishing feature is that, rather than being locked to $\left<111\right>$, the low-energy ground states of bulk BiFeO\ts{3}, $\vec{u}(\vec{r})$ rotates continuously in our BiFeO\ts{3} layers. The realization of a continuous $U(1)$ Lie group is consistent with a vortex-sustaining order-parameter manifold $\mathcal{M}$ and results from an effective suppression of crystalline anisotropy by competing depolarization and dipole fields emerging with the polar discontinuity at the multiferroic/dielectric interface. Indeed,  Figs. \ref{fig:TEM}(a)-(c) establish an ordered vortex phase in which 1-nm-wide vortex cores with positive topological charge appear staggered above and below the centerline of each BiFeO\ts{3} layer. In analogy to the swirl of tropical storms located above and below the equator, only vortices with positive (negative) chirality are observed above (below) the BiFeO\ts{3} centerline. We call this novel vortex phase $\pi_1$-$S_\infty$, wherein $\pi_1$ reflects a nontrivial fundamental group and $S_\infty$ denotes the frieze group of the order parameter morphology. % In-plane core separations are consistent with XRD and DF-TEM findings [Figs. \ref{fig:XRD}(d), \ref{fig:TEM}(a), and \ref{fig:TEM}(b)].

%\section{Phase-field Results}

\begin{figure}[!hbt]
\includegraphics[width=0.47\textwidth]{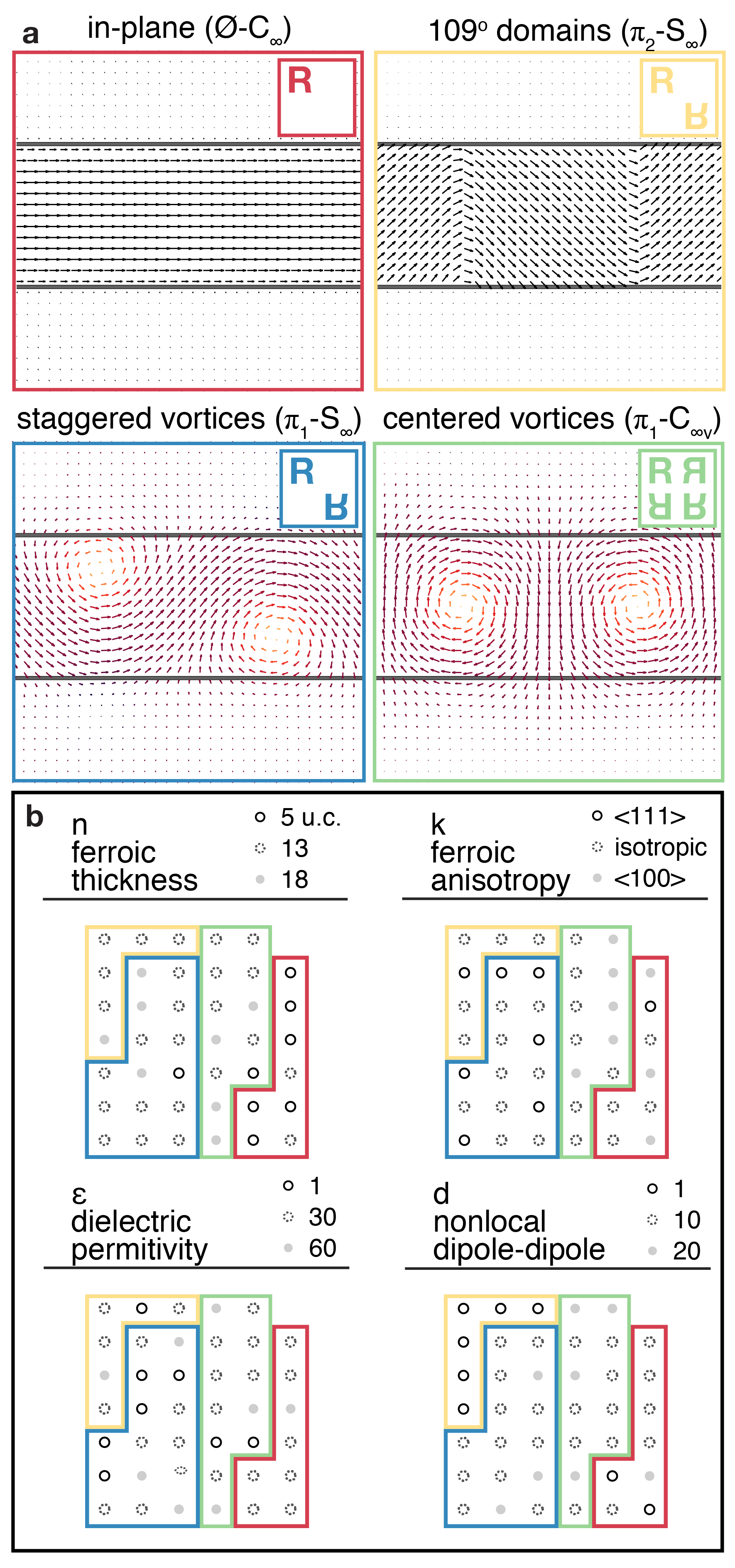}
\caption{\label{fig:SIM}
\textbf{Simulated order-parameter morphologies.}
(a) Representative morphologies $\vec{u}(\vec{r})$ of four competing phases ($\emptyset$-$C_\infty$, $\pi_2$-$S_\infty$, $\pi_1$-$S_\infty$, and $\pi_1$-$C_{\infty v}$). Frieze symmetry operations which leave $\vec{u}(\vec{r})$ invariant (inserts).
(b) Phase diagram showing the generating set of $\{n,k,d,\epsilon\}$ parameters engendering the phases in (a).  $n$ is the ferroic layer thickness, $k$ is the order-parameter anisotropy, $\epsilon$ is the dielectric permittivity, and $d$ is the a coefficient regulating the strength of nonlocal dipole-dipole interactions.
}
\end{figure}

The morphology of the $\pi_1$-$S_\infty$ ordered vortex phase can be understood by developing a field model based on a free energy functional 
\begin{align}
\label{eq:A}
	\mathcal{L} \left[ \vec{u}(\vec{r}) \right] = \int 
		\mathcal{G}[\vec{u}(\vec{r})] +
		\mathcal{D}[\vec{u}(\vec{r})] +
		\mathcal{U}[\vec{u}(\vec{r})]
	d\vec{r},
\end{align}
comprised of three essential terms,
\begin{subequations}
\label{eq:model}
\begin{align}
	\mathcal{G}[\vec{u}(\vec{r})] &= \left|\vec{\nabla}_\vec{r} \vec{u}(\vec{r}) \right|^2 \\
	\mathcal{D}[\vec{u}(\vec{r})] &= \vec{u}(\vec{r}) \cdot \int (\vec{\nabla}_\vec{r}\otimes\vec{\nabla}_{\pvec{r}'})G(\vec{r}-\pvec{r}')\vec{u}(\pvec{r}') d\pvec{r}' \\
	\mathcal{U}[\vec{u}(\vec{r})] &= - \frac{1}{2} |\vec{u}(\vec{r})|^2 + \frac{1}{4} |\vec{u}(\vec{r})|^4 .
\end{align}
\end{subequations}
$\mathcal{G}[\vec{u}(\vec{r})]$ represents the order parameter stiffness by associating an energetic cost with the magnitude of the field gradient $\left|\vec{\nabla}_\vec{r} \vec{u}(\vec{r}) \right|$; $\mathcal{D}[\vec{u}(\vec{r})]$ is a non-local dipole-dipole interaction which acts, through $G(\vec{r}-\pvec{r}')$ the Green function of the Laplacian, to suppress stray $\vec{u}(\vec{r})$ fields;\cite{Abert:2012go,Maugin:2013wl} and $\mathcal{U}[\vec{u}(\vec{r})]$ is the potential energy function of the order parameter. For isotropic dielectrics, $\mathcal{U}[\vec{u}(\vec{r})]$ is harmonic with a proportionality coefficient equal to $\epsilon$ the dielectric constant, i.e. $\epsilon |\vec{u}(\vec{r})|^2 / 2$. The anharmonic potential in Eq. \ref{eq:model}(c) embodies an idealized ferroic material for which the order parameter has condensed into a degenerate ground state isomorphic with $U(1)$. Crystalline anisotropies are incorporated via multipole expansions.\cite{Jackson:2007ub}

Unsupervised machine learning\cite{VanDerMaaten:2008tm,Anderberg:2014wx,Fried:2015kx} is employed to classify $\vec{u}(\vec{r})$  ground states, obtained by relaxing Eq. \ref{eq:A}, based on topological structure and frieze symmetry.\cite{Kopsky:2010hv} Representative $\vec{u}(\vec{r})$ patterns and corresponding generating parameter sets are presented in Fig. \ref{fig:SIM}(a) and \ref{fig:SIM}(b), respectively. Morphologies identified as $\emptyset$-$C_\infty$ represent a topologically trivial phase consisting of an order parameter field that is fully aligned along the ferroic lamella. Formation of this phase is encouraged by small ferroic layer thicknesses $n$ and dielectric permittivities $\epsilon$ -- two factors which contribute to suppressing out-of-plane order-parameter components. The pairing of thicker $n$ with small dipole-dipole interactions $d$ produce domain walls -- the defining characteristic of $\pi_2$-$S_\infty$ a phase which is commonly observed in monolithic BiFeO\ts{3} layers.\cite{Chen:2015cg}

Enhancing dipole-dipole interactions encourages the formation of two competing vortex phases, $\pi_1$-$S_\infty$ and $\pi_1$-$C_{\infty v}$. For $\pi_1$-$S_\infty$, vortex cores are staggered based on chirality, giving rise to an order-parameter morphology agreeing with $\vec{u}(\vec{r})$ observed in our coherent multiferroic BiFeO\ts{3}-based superlattices. For $\pi_1$-$C_{\infty v}$, vortex cores are centered, reproducing the pattern observed in ferroelectric PbTiO\ts{3}-based superlattices.\cite{Yadav:2016ju} When the order-parameter potential is isotropic, both phases are stable. The introduction of weak anisotropies $k$ canting $\vec{u}(\vec{r})$ towards $\left<100\right>$ however suppress $\pi_1$-$S_\infty$, the phase characterized by chiral staggering, in favor of $\pi_1$-$C_{\infty v}$, the phase with centered vortices.

%\section{Conclusion}

In summary, the effective order-parameter potential energy landscape of BiFeO\ts{3} is engineered through a combination of symmetry templating and dimensional confinement in order to produce an order-parameter manifold $\mathcal{M}$ obeying $U(1)$ Lie group symmetries, the simplest group exhibiting a nontrivial fundamental homotopy group capable of supporting vortices. Direct and Fourier-space analyses are combined to establish the realization of an ordered vortex phase $\pi_1$-$S_\infty$ with cores characterized by positive topological charge and chiral staggering in coherent magnetoelectric multiferroic TbScO\ts{3})$_{m}$/(BiFeO\ts{3})$_{n}$ superlattices. Using a field model describing the local order-parameter stiffness and competing non-local dipole-dipole interactions, the experimentally observed order-parameter morphology is reproduced. Anisotropies canting the order parameter towards $\left<100\right>$ is found to suppress chiral staggering and center the vortices.

\section*{Acknowledgements}

This research was supported by the Army Research Office under grant W911NF-16-1-0315.  This work made use of the Cornell Center for Materials Research (CCMR) Shared Facilities, which are supported through the NSF MRSEC program (No. DMR-1719875).  Substrate preparation was performed in part at the Cornell NanoScale Facility, a member of the National Nanotechnology Coordinated Infrastructure (NNCI), which is supported by the NSF (Grant No. ECCS-1542081).

\vfill\clearpage

%\bibliography{../Bibliography.bib}


\end{document}